\documentclass[aps,showpacs]{revtex4}
\usepackage{graphicx}
\usepackage{amsmath}
\usepackage{amssymb}
\newcommand{\be}{\begin{equation}}
\newcommand{\ee}{\end{equation}}
\newcommand{\ben}{\begin{eqnarray}}
\newcommand{\een}{\end{eqnarray}}
\newcommand{\bes}{\begin{subequations}}
\newcommand{\ees}{\end{subequations}}
\newcommand{\bb}{\bibitem}
\newcommand{\sech}{{\rm sech}}

\begin{document}
\title{New braneworld models in the presence of auxiliary fields}
\author{D. Bazeia$^{1}$\footnote{Corresponding author. Email: bazeia@fisica.ufpb.br}, M.A. Marques$^1$, R. Menezes$^{2,3}$, and D.C. Moreira$^1$}
\affiliation{$^1$Departamento de F\'\i sica, Universidade Federal da Para\'\i ba, 58051-970 Jo\~ao Pessoa, PB, Brazil}
\affiliation{$^2$Departamento de Ci\^encias Exatas, Universidade Federal da Para\'{\i}ba, 58297-000 Rio Tinto, PB, Brazil}
\affiliation{$^3$Departamento de F\'\i sica, Universidade Federal de Campina Grande, 58109-970, Campina Grande, PB, Brazil}

\begin{abstract}
We study braneworld models in the presence of auxiliary fields. We use the first-order framework to investigate several distinct possibilities, where the standard braneworld scenario changes under the presence of the parameter that controls the auxiliary fields introduced to modify Einstein's equation. The results add to previous ones, to show that the minimal modification that we investigate contributes to change quantitatively the thick braneworld profile, although no new qualitative effect is capable of being induced by the minimal modification here considered.
\end{abstract}

\date{\today}


\maketitle

\section{Introduction}

In this work we deal with braneworld models in the presence of scalar fields \cite{RS,GW,F,C,AH,KOP,soda}, going  beyond General Relativity (GR) with the addition of auxiliary fields. As one knows, one important problem in the construction of alternative theories of gravity with the addition of extra dynamical fields is the presence of extra degrees of freedom which in general lead to instabilities in these theories \cite{ghost1,ghost2}. However, an interesting way to circumvent this problem was suggested very recently in Ref.~\cite{psv}, in which one uses non-dynamical or auxiliary fields. 

The modification introduced in the presence of auxiliary fields has then been studied in Refs.~\cite{c1,c2} within the cosmological context, to see how the auxiliary fields may contribute to the cosmic evolution. Moreover, it has also been recently studied within the thick braneworld context, in five dimensions with a single extra dimension of infinite extent \cite{gly,we}. In \cite{gly}, the authors investigated the problem numerically, and found that the braneworld scenario in the presence of auxiliary field remains linearly stable, and in \cite{we} one investigates the case with a simpler extra term, controlled by a single real parameter which indicates deviation from GR. There, one introduced a first-order framework, which helps to find analytical solutions for both the scalar field and warp factor, and two distinct models were studied. 

These investigations motivate us to further study the thick braneworld scenario in the presence of auxiliary fields, but now extending the first-order framework introduced in \cite{we} to scalar field models described by one and two real scalar fields, searching for exact solutions in a braneworld scenario with a single extra spatial dimension of infinite extent. We consider the case which is controlled by a bulk scalar field, leading to the thick brane scenario that is well explained in Refs.~\cite{F,C,soda}, so we omit the details in the current work. In order to implement the investigation, in Sec.~\ref{sec:gen} we generalize the results of Ref.~\cite{we} to the case of several scalar fields. We then study new models in Sec.~\ref{sec:models}, where we deal with the splitting of the solution \cite{p}, the presence of a second scalar field, leading to the Bloch brane scenario \cite{bg}, and with another interesting possibility, leading to the construction of a hybrid brane scenario \cite{blmm}. We end the work in Sec.~\ref{sec:end}, where we add some comments and conclusions.

\section{Generalities}
\label{sec:gen}

Let us start following Refs.~\cite{psv,we}. We consider the following modified Einstein equations with auxiliary fields:
\be
G_{ab} = 2T_{ab} + S_{ab},
\ee
where 
\be\label{originaleq}
S_{ab}=\alpha_1 g_{ab} T + \alpha_2 g_{ab}T^2 + \alpha_3 T T_{ab}+ \alpha_4 g_{ab}T_{cd}T^{cd} + \alpha_5 T_a^cT_{cb} + \beta_1\nabla_a\nabla_b T + \beta_2 g_{ab}\Box T + \beta_3\Box T_{ab} + 2\beta_4\nabla^c\nabla_{(a}T_{b)c} + \ldots
\ee
However, we keep only non-derivate linear terms in $T$ and assume $S_{ab}<<T_{ab}$ in order to maintain the above modified Einstein equations divergent free with a non-dynamical field parametrized by a real parameter $\alpha$ as follows
\be\label{einsteineq}
R_{ab} - \frac12 g_{ab}R = 2T_{ab} + \alpha g_{ab}T,
\ee
where we have changed $\alpha_1\to\alpha$. Moreover, we use $a,b=0,1,..,4$, $4\pi G=1$, $T=g_{ab}T^{ab}$. In the original work \cite{psv}, the modification includes auxiliary fields up to fourth-order in derivatives, as we can see in Eq.~\eqref{originaleq}. For this reason, we refer to the above expression \eqref{einsteineq} as the minimal modification. This is of interest, since the minimal modification allows that we write a first-order framework which helps us to obtain analytical solutions, simplifying the investigation of stability of the gravity sector, as already advanced in Ref.~\cite{we}. 

We turn attention to the braneworld scenario, recalling that the line element for a thick brane model in a 5-dimensional spacetime with a single extra dimension of infinite extent is
\be\label{lineelement}
ds^2=g_{ab}dx^a dx^b = e^{2A}\eta_{\mu\nu}dx^\mu dx^\nu - dy^2,
\ee
with $\eta_{\mu\nu}={\rm diag}(+,-,-,-)$ representing the metric of the 4-dimensional Minkowski spacetime. $A$ is the warp function and $e^{2A}$ is the warp factor.

To source the thick braneworld model, we take the standard Lagrange density for the set of $k=1,2,...,N$ real scalar fields
\be
{\cal L} = \frac12 \sum_k\partial_a \phi_k\partial^a \phi_k - V(\phi_1,\ldots,\phi_N).
\ee
In the braneworld scenario, both the warp function and scalar fields only depend on the extra dimension, that is, $A=A(y)$ and
$\phi_k=\phi_k(y)$. In this case, the energy-momentum tensor assumes the form
\be
T_{ab}=g_{ab}\left(\frac12 \sum_k\phi_k^\prime\phi_k^\prime + V \right) + \sum_k\partial_a\phi_k \partial_b\phi_k,
\ee
with the trace $T$ given by 
\be
T=\frac32 \sum_k \phi_k^\prime\phi_k^\prime + 5V.
\ee
Here prime means derivative with respect to the extra dimension $y$.
The nonzero components of Eq.~\eqref{einsteineq} for the line element \eqref{lineelement} are then given by
\bes
\be\label{eqeinsteina}
6{A^\prime}^2=\frac{2-3\alpha}{2}\sum_k \phi_k^\prime\phi_k^\prime-(2+5\alpha)V,
\ee
and
\be\label{eqeinsteinb}
A^{\prime\prime}=-\frac23\sum_k \phi_k^\prime\phi_k^\prime.
\ee
\ees
This last equation does not depend on $\alpha$, so the modification included in Eq.~\eqref{einsteineq} does not change its behavior. Now, we introduce the superpotential $W=W(\phi_1,\ldots,\phi_N)$ to write the first-order equations
\bes
\be\label{firstorderb}
\phi_k^\prime=\frac12 W_{\phi_k},
\ee
and
\be\label{firstordera}
A^\prime=-\frac13 W,
\ee
\ees
where $W_{\phi_k}=\partial W/\partial\phi_k$. In this case, the potential has to be
\be\label{potential}
V=\frac{2-3\alpha}{8(2+5\alpha)} \sum_k W_{\phi_k}W_{\phi_k}-\frac{2}{3(2+5\alpha)} W^2,
\ee
and so we suppose that the parameter $\alpha$ varies inside the open interval $(-2/5,2/3)$. 

In the presence of $W$, the energy density can be written as 
\be\label{rho}
\rho(y)=\frac{e^{2A}}{2+5\alpha}\left(\frac{2+\alpha}{4}\sum_k W_{\phi_k}W_{\phi_k}-\frac{2}{3}\,W^2\right).
\ee
As a result obtained before in Ref.~\cite{we}, here we also note that the first-order Eqs.~\eqref{firstorderb} and \eqref{firstordera}
do not depend on $\alpha$, so the solutions are the same of the standard braneworld model. However, both the potential and energy density
change with $\alpha$, and control the way one deviates from the standard scenario.

Another issue of current interest concerns stability of the gravity sector in the presence of several scalar fields.
However, from \cite{we} we note that the stability behavior in the gravity sector only depends on the warp factor, and since it does not change with $\alpha$, the presence of auxiliary fields does not modify linear stability. Thus, the gravity sector in the presence of several scalar fields is also robust, as it is in the standard scenario, for $\alpha=0$.

The above results generalize the previous results \cite{we} to the case of several real scalar fields. This is of interest since we can now investigate several models, to see how the $\alpha$ parameter works to change the thick braneworld scenario. To make this specific, in the next section we investigate three distinct models. 

\section{New models}
\label{sec:models}

Let us now study some models described by one and two real scalar fields, with the motivation to understand how the modification controlled by
$\alpha$ contributes to change the standard braneworld scenario.

\subsection{The p-brane model}

\begin{figure}[t!]
\includegraphics[width=5cm]{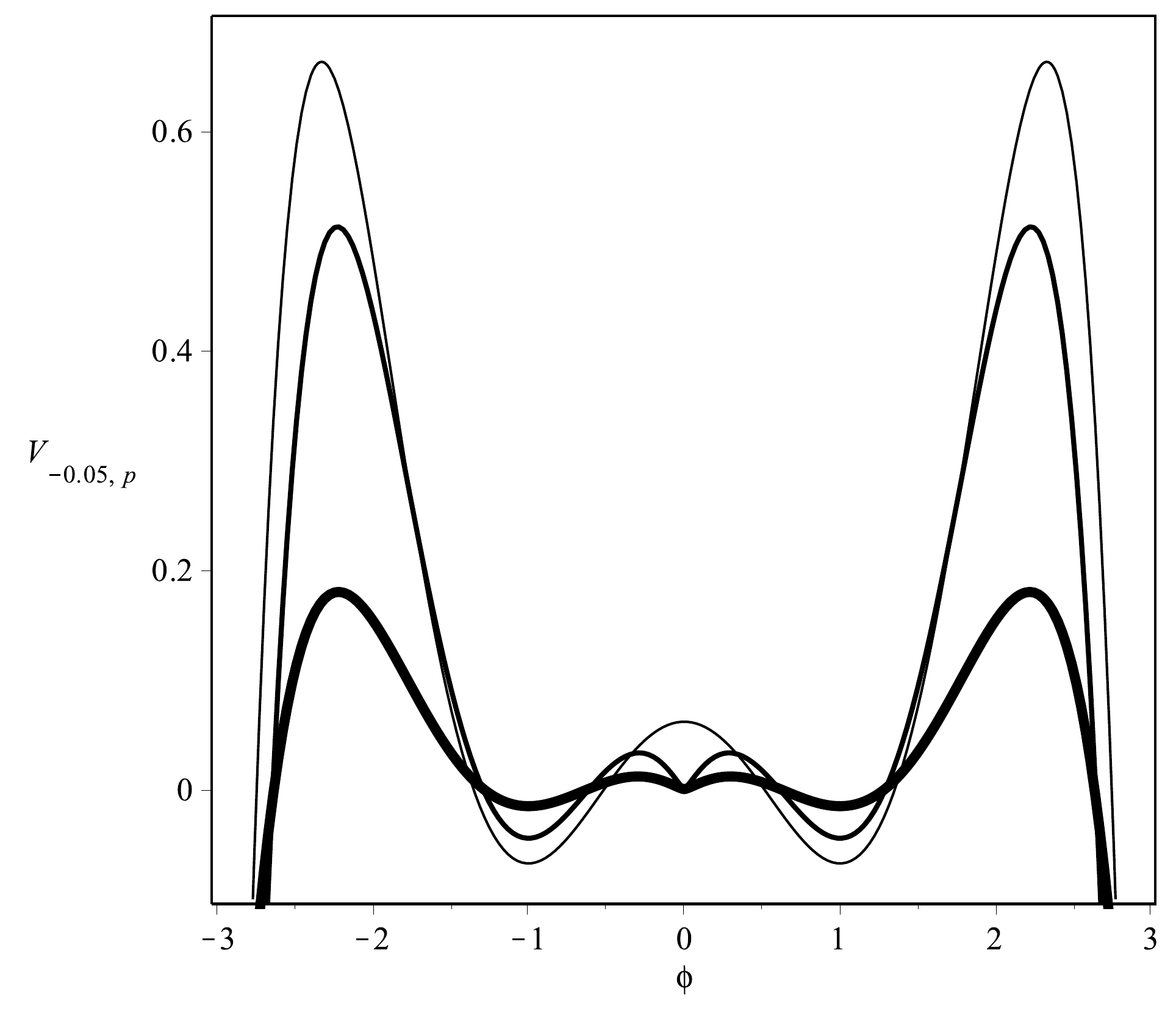}
\hspace{1cm}
\includegraphics[width=5cm]{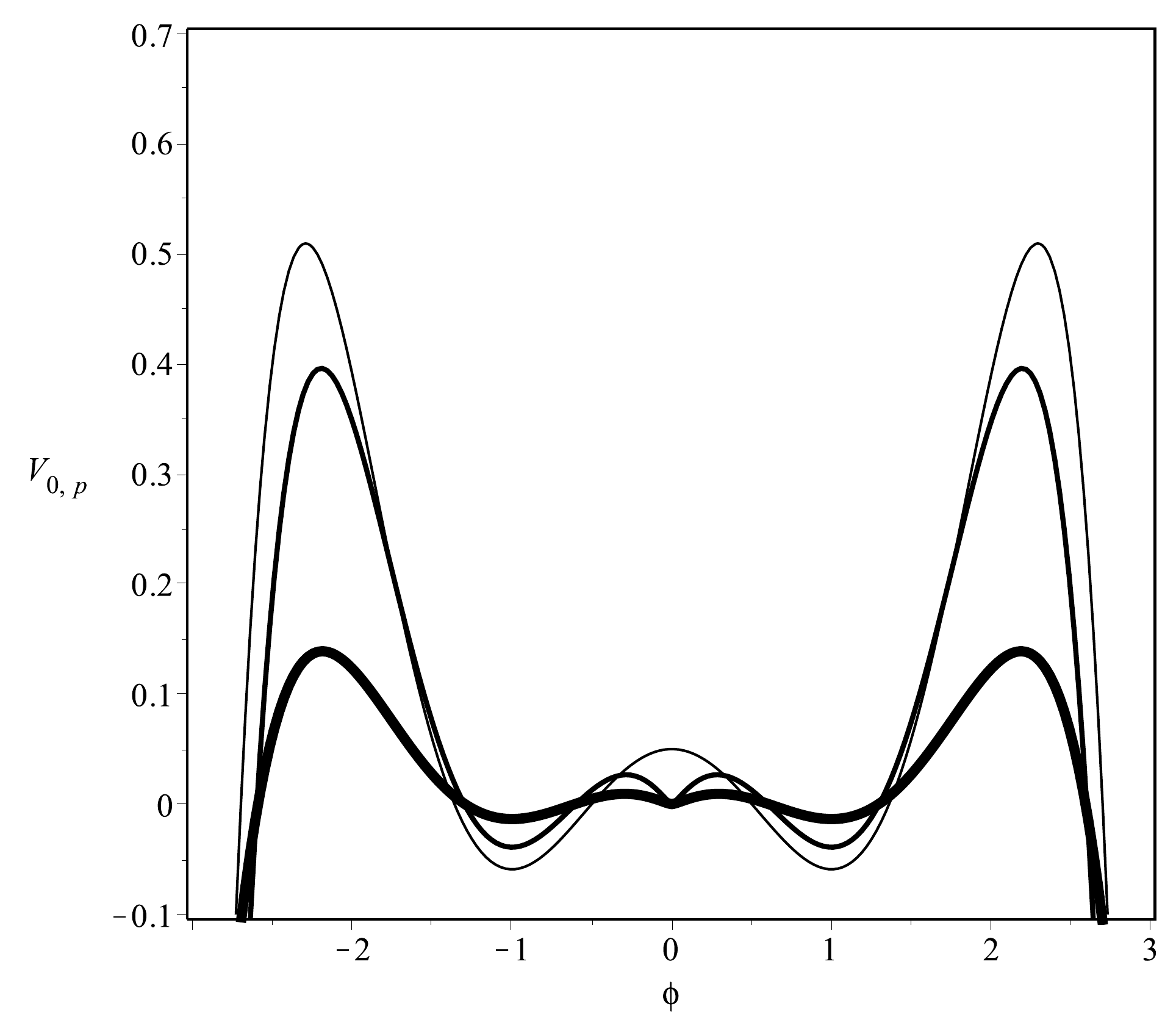}
\hspace{1cm}
\includegraphics[width=4.7cm]{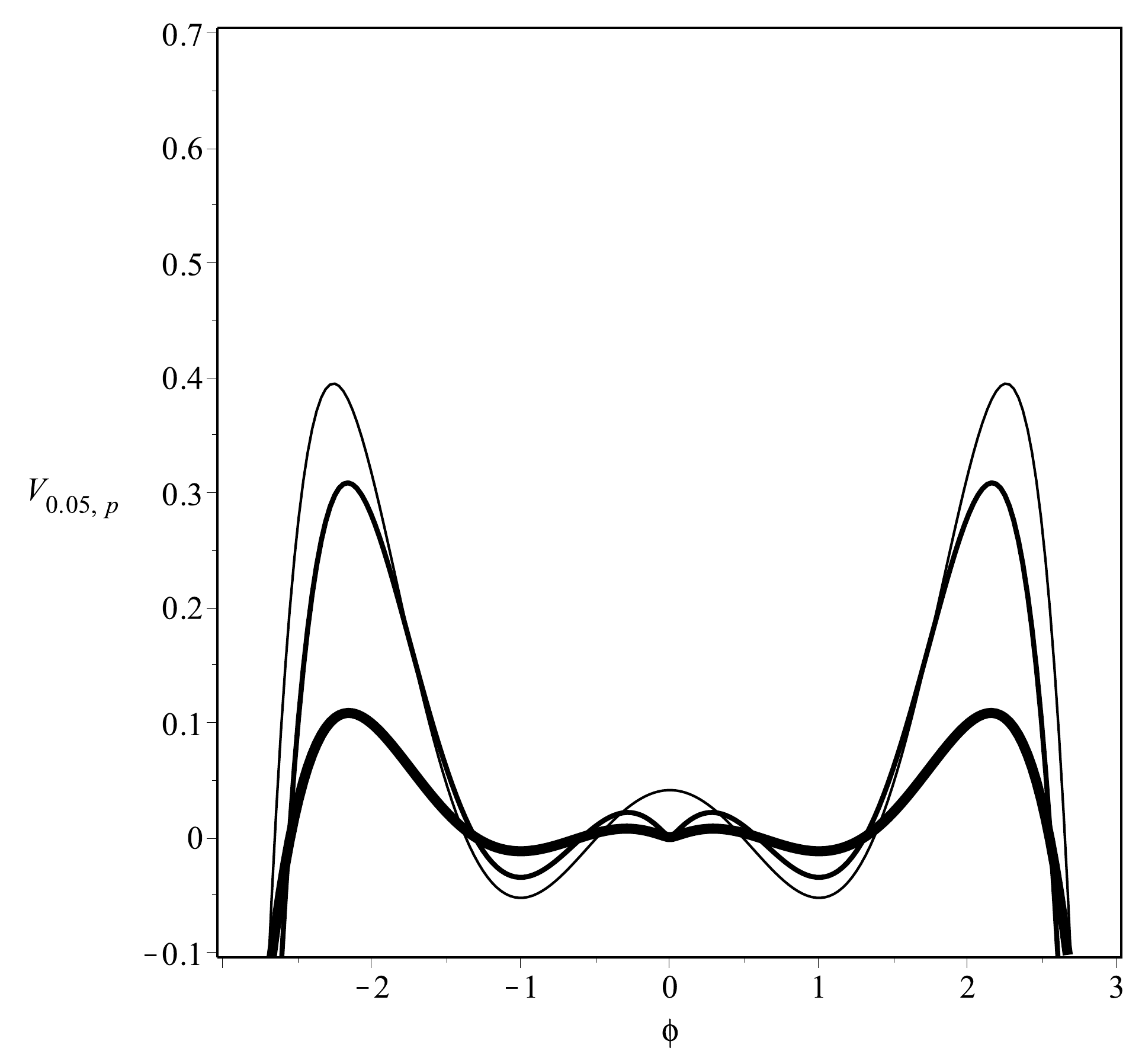}
\caption{The potential $V_{\alpha,p}$ plotted for $\alpha=-0.05$ (left), $\alpha=0$ (center) and $\alpha=0.05$ (right). To ease comparison, we have plotted 1/10 of $V_{\alpha,1}$ in each figure. Also, we used $p=1, 3$ and $5$, with the thickness of the lines growing with increasing $p$.}
\label{fig1}
\end{figure}

We consider the model first investigated in \cite{p}. It is described by a single real scalar field, and has $W=W(\phi)$ given by 
\be\label{pmodel}
W_p(\phi)=\frac{2p}{2p-1}\phi^{\frac{2p-1}{p}} - \frac{2p}{2p+1}\phi^{\frac{2p+1}{p}},
\ee
in which $p$ is odd integer, $p=1,3,5,...$\,, otherwise the above function can admit imaginary values. This model is of interest because it may induce the splitting of the brane, a feature that appears from the energy density. In the standard thick brane scenario, the energy density presents a maximum at the origin, inside the brane, but in the model \eqref{pmodel} it vanishes at the origin, splitting the brane, as we illustrate below. In the braneworld model under investigation, the potential becomes
\be
V_{\alpha,p}(\phi)=\frac{2-3\alpha}{2(2+5\alpha)}\left(\phi^{\frac{p-1}{p}}- \phi^{\frac{p+1}{p}}\right)^2-\frac{8}{3(2+5\alpha)}\!\!\left(\!\frac{p}{2p-1}\phi^{\frac{2p-1}{p}}\! \!\!-\! \frac{p}{2p+1}\phi^{\frac{2p+1}{p}}\! \right)^2.
\ee
It is plotted in Fig.~\ref{fig1}, for several values of $\alpha$ and $p$. We see that both $p$ and $\alpha$ contribute to control the amplitude of the extrema of the potentials, with the modifications being less sensitive to $\alpha$.

\begin{figure}[t!]
\includegraphics[width=6cm]{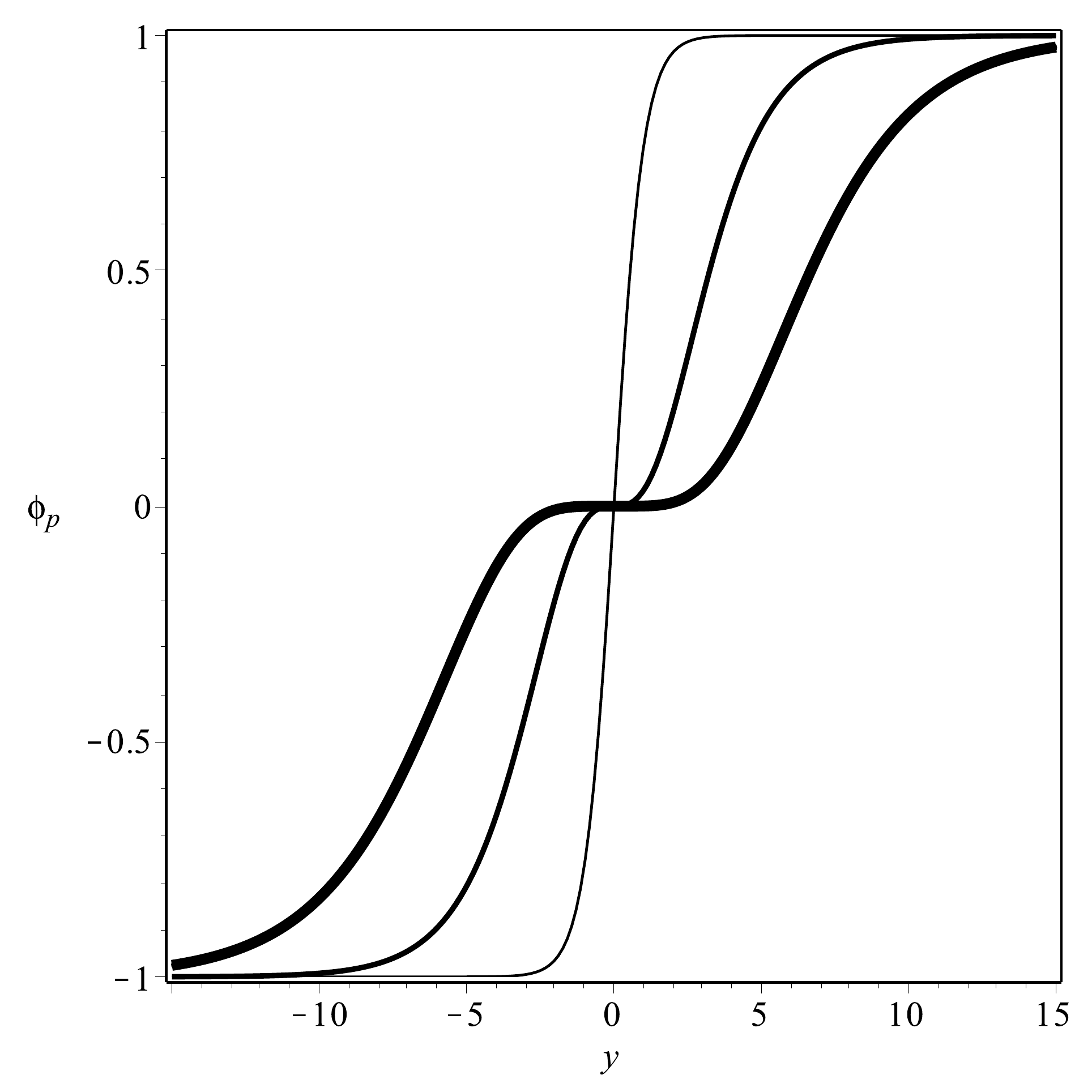}
\caption{The solution \eqref{solphip} for $p=1, 3$ and $5$, with the thickness of the lines growing with increasing $p$.}
\label{fig2}
\end{figure}

In this case, the equation \eqref{firstorderb} becomes
\be
\phi^\prime = \phi^{\frac{p-1}{p}} - \phi^{\frac{p+1}{p}},
\ee
and its solution is
\be\label{solphip}
\phi_p(y) = \tanh^p\left(\frac yp\right).
\ee
In Fig.~\ref{fig2} we depict the behavior of $\phi_p$ with $p$. The width of the solution increases when $p$ increases.

\begin{figure}[htb!]
\includegraphics[width=6cm]{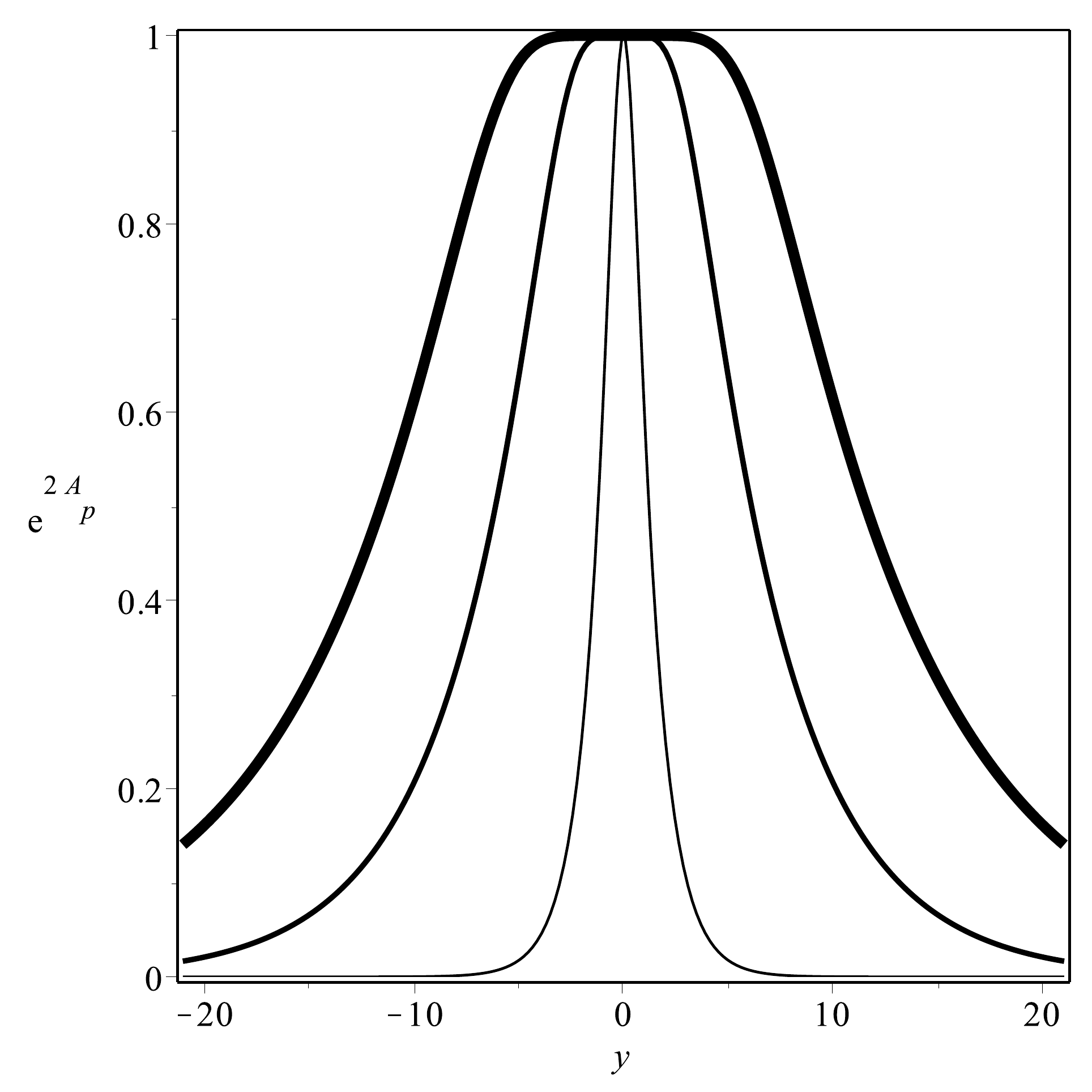}
\caption{The warp factor for the warp function \eqref{warpfunctionp} depicted for $p=1, 3$ and $5$, with the thickness of the lines growing with increasing $p$.}
\label{fig3}
\end{figure}

We can use Eq.~\eqref{firstordera} to obtain the warp function
\be\label{warpfunctionp}
A_p(y)\!=\! -\frac13\!\! \left(\frac{p}{2p+1} \right)\!\tanh^{2p}\!\!\left(\frac{y}{p} \right) \!- \frac23\!\!\left(\!\frac{p^2}{2p-1}-\frac{p^2}{2p+1}\!\right)\left(\ln\cosh\left(\frac{y}{p} \right) -\sum_{n=1}^{p-1}\frac{1}{2n} \tanh^{2n}\left(\frac{y}{p} \right)\right),
\ee
which we use to depict the warp factor in Fig.~\ref{fig3}.

\begin{figure}[htb!]
\includegraphics[width=5.8cm]{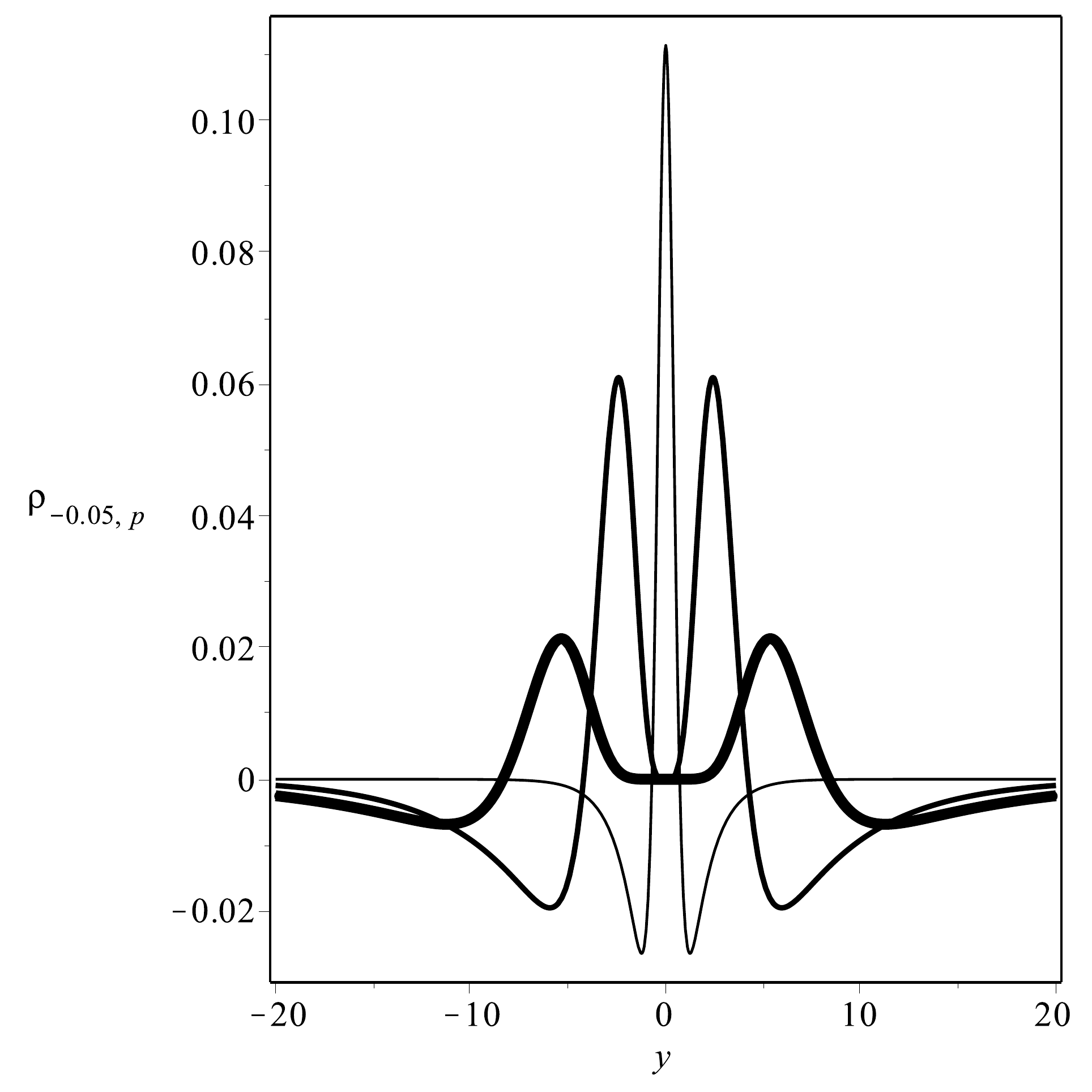}
\includegraphics[width=5.8cm]{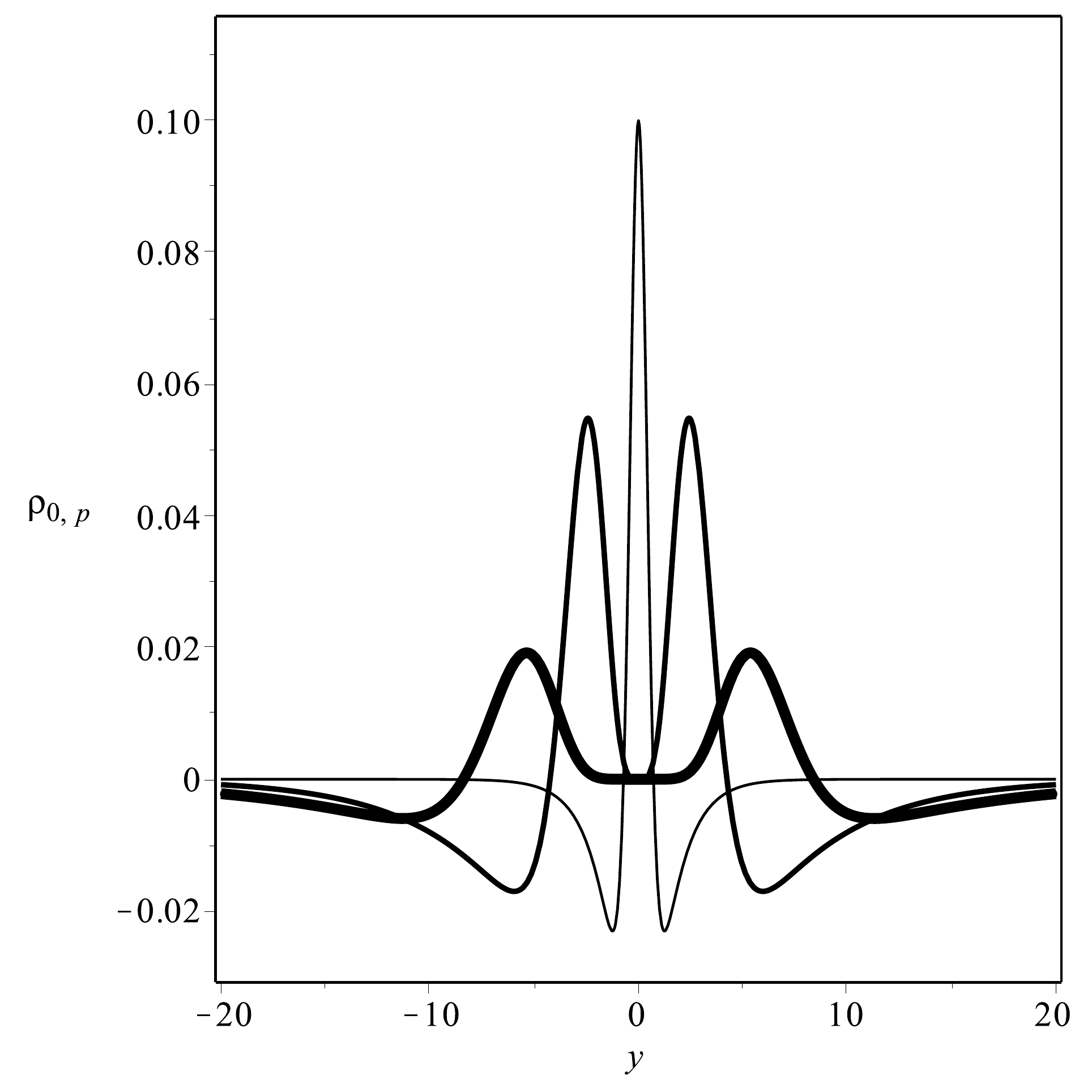}
\includegraphics[width=5.8cm]{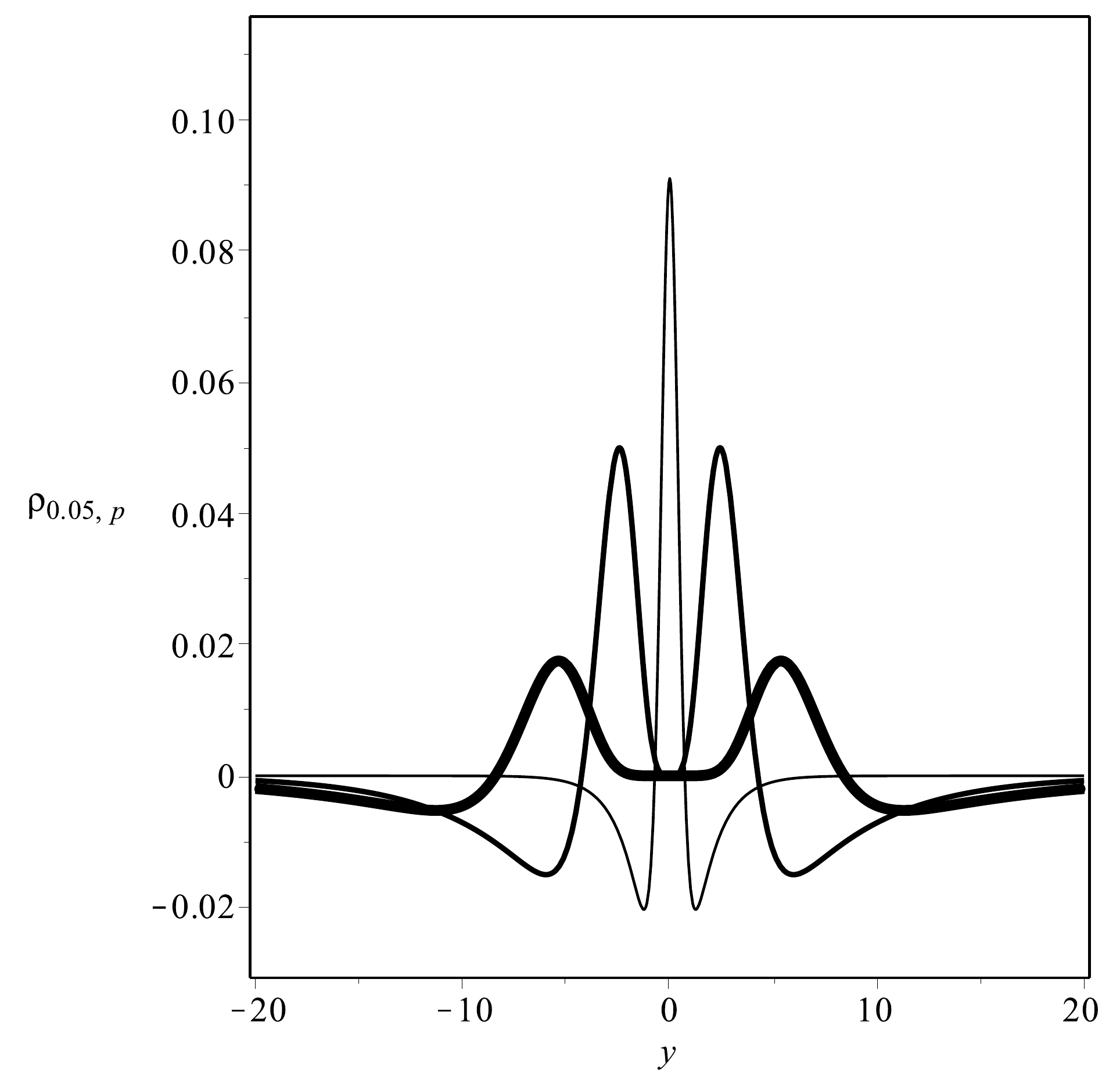}
\caption{The energy density $\rho_{\alpha,p}$ depicted for $\alpha=-0.05$ (left), $\alpha=0$ (center) and $\alpha=0.05$ (right). We plotted 1/10 of $\rho_{\alpha,1}$ in each figure, and we used $p=1, 3$ and $5$. with the thickness of the lines growing with increasing $p$.}
\label{fig4}
\end{figure}

The analytic expression for the energy density $\rho_{\alpha,p}$ is awkward, so we depict it in Fig.~\ref{fig4} to show how it responds to the parameters $\alpha$ and $p$. We see from Fig.~\ref{fig4} that the variation of $\alpha$ does not modify qualitatively the energy density, although there is some quantitative modification induced by the auxiliary fields. In particular, we note that for $p=3,5$, fixed, the increasing of $\alpha$ contributes to change the height of the maxima and the depth of the minima. However, it does not change significantly the region with zero energy density inside the brane, so it does not contribute to amplify the splitting of the brane.

\subsection{The Bloch brane model}

In this particular case, we work with two scalar fields, $\phi$ and $\chi$. The model was first studied in a thick braneworld scenario in \cite{bg}. The superpotential for this model is
\be\label{superpotentialbnrt}
W_r(\phi,\chi)=2\phi-\frac23 \phi^3-2r\phi\chi^2,
\ee
where $r$ is a positive real parameter. The minima are at the points $(\pm1,0)$ and $(0,\pm 1/\sqrt{r})$. By using Eq.~\eqref{potential}, we find the potential
\be
V_{\alpha,r}(\phi,\chi)=\frac{2-3\alpha}{2(2+5\alpha)}\left((1-\phi^2-r\chi^2)^2+4r^2\phi^2\chi^2\right)-\frac{8}{3(2+5\alpha)}\left(\phi-\frac{\phi^3}{3}-r\phi\chi^2\right)^2.
\ee
This potential is a surface in the $(\phi,\chi)$ plane. It is not easy to see how it changes when $\alpha$ and $r$ vary, so we do not depict it here.

\begin{figure}[htb!]
\includegraphics[width=4.2cm]{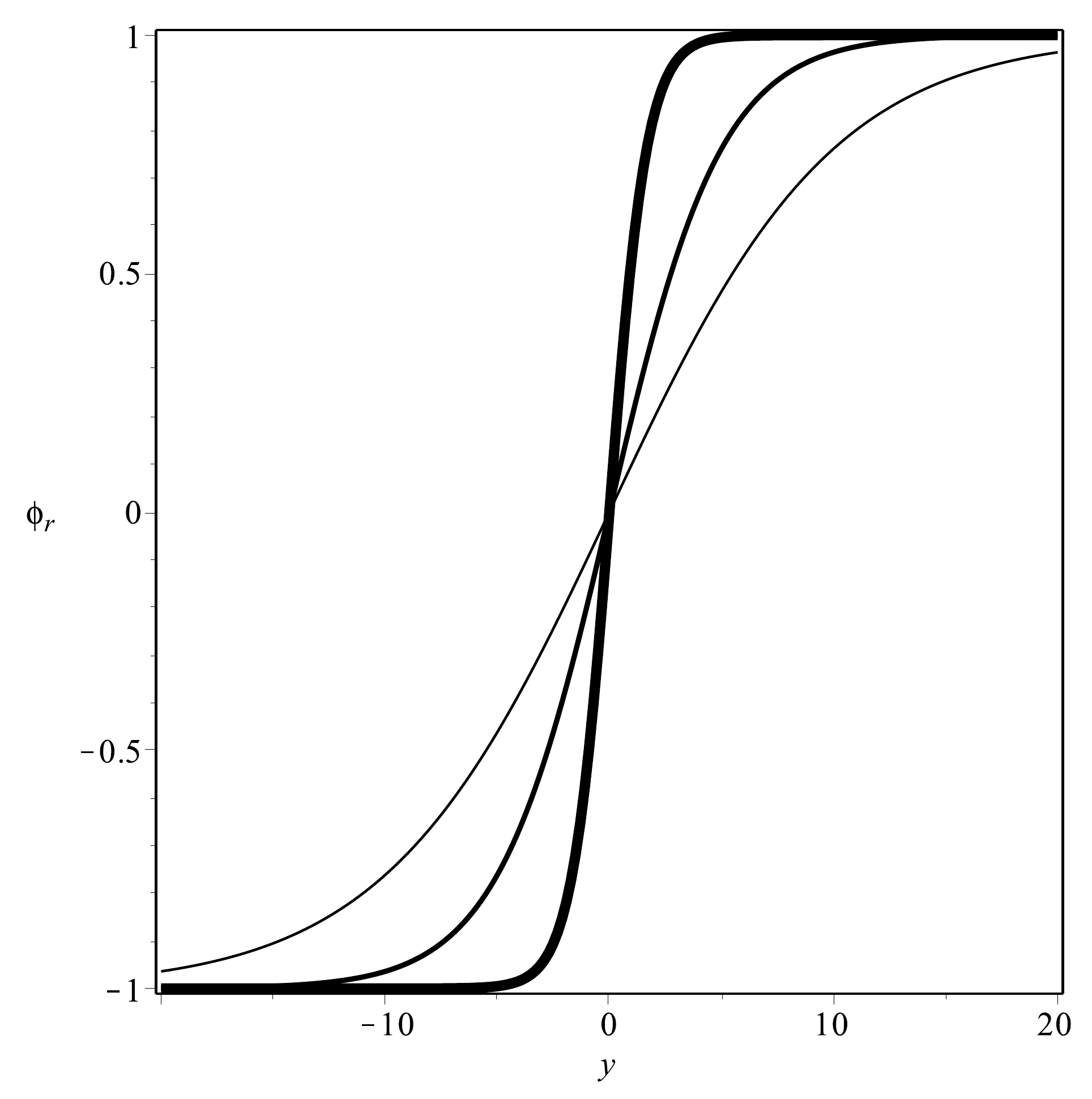}
\includegraphics[width=4.2cm]{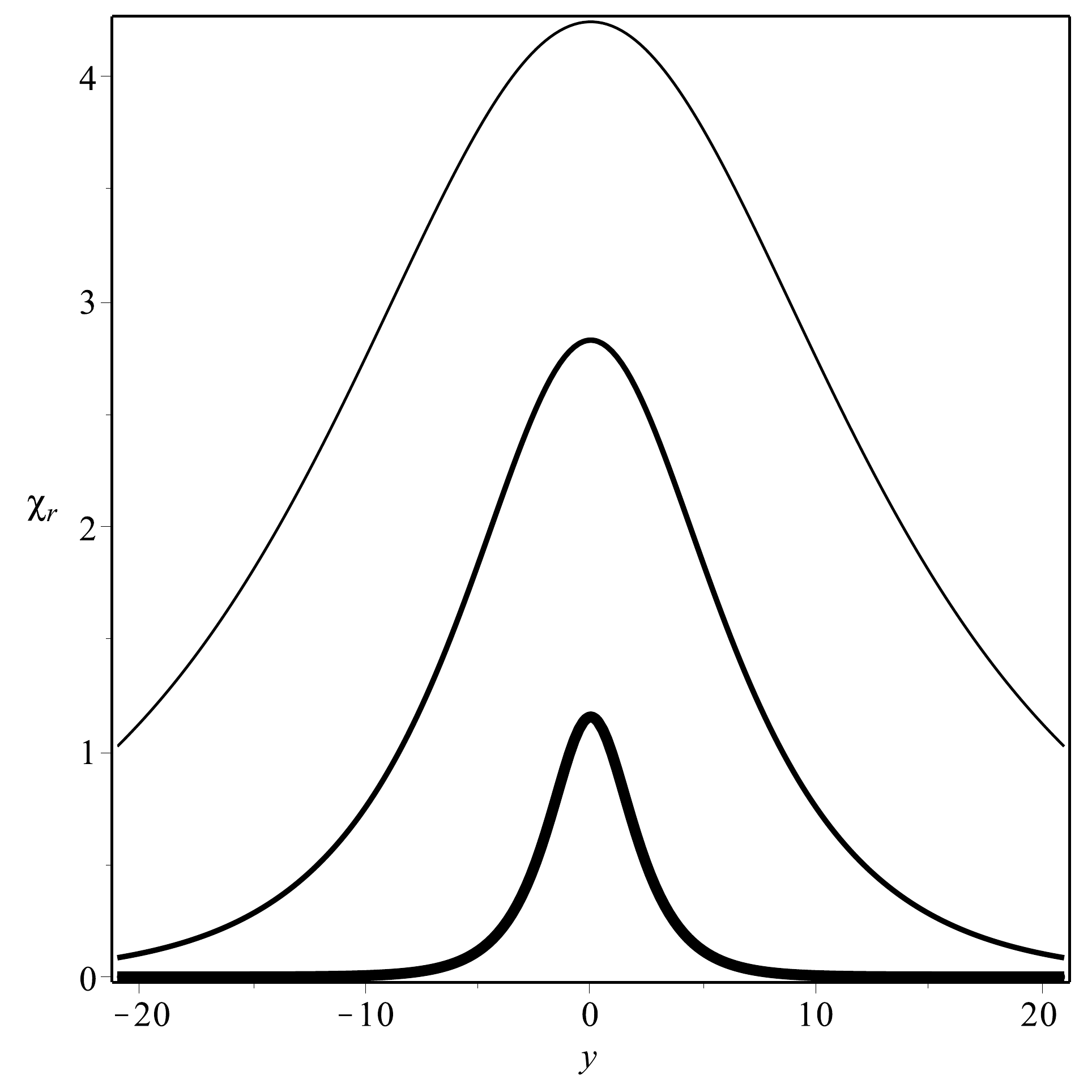}
\caption{The solutions \eqref{solphir} and \eqref{solchir}, respectively, for $r=0.05, 0.1$ and $0.3$. The thickness of the lines grows as
$r$ increases.}
\label{fig5}
\end{figure}

The equations \eqref{firstorderb} and \eqref{firstordera} give
\bes
\be
\phi^\prime = 1-\phi^2-r\chi^2,
\ee
and
\be
\chi^\prime=-2r\phi\chi.
\ee
\ees
We take the eliptic orbit, for $r\in (0,1/2)$,
\be
\phi^2+\frac{r}{1-2r}\chi^2=1.
\ee
Now the solutions connecting the minima $(\pm1,0)$ are
\bes
\be\label{solphir}
\phi_r(y)=\tanh(2ry),
\ee
and
\be\label{solchir}
\chi_r(y)=\pm\sqrt{\frac1r-2}\;\sech(2ry).
\ee
\ees
They are plotted in Fig.~\ref{fig5}.

\begin{figure}[htb!]
\includegraphics[width=6cm]{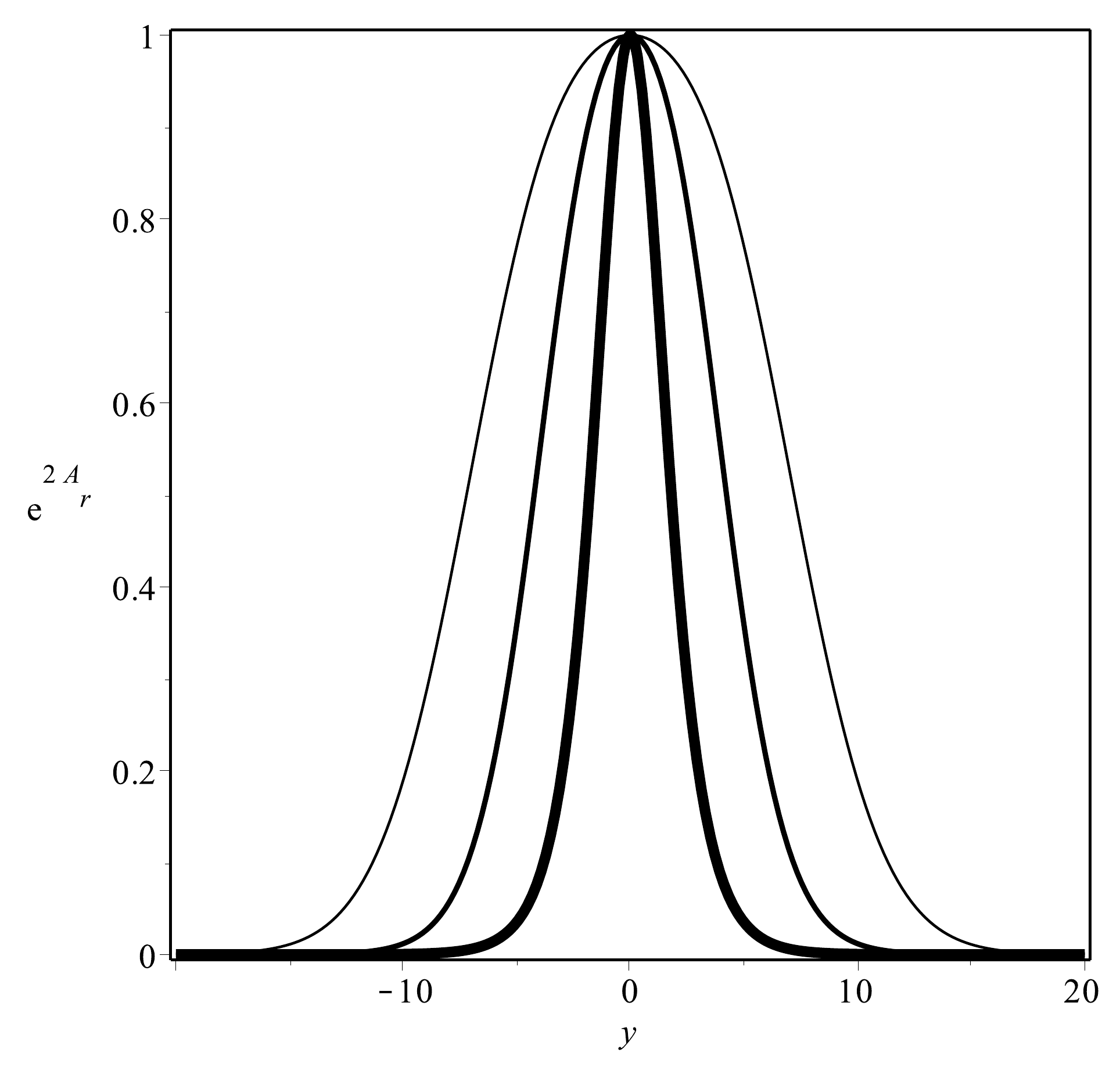}
\caption{The warp factor for the warp function \eqref{warpfunctionr} for $r=0.05, 0.1$ and $0.3$. The thickness of the lines grows as
$r$ increases.}
\label{fig6}
\end{figure}

We notice that in the limit $r\to1/2$, the above two-field solution changes to the solution with $\phi(y)=\tanh(y)$ and $\chi(y)=0$. By using the two-field solution, it is possible to solve the Eq.~\eqref{firstordera} to get to the warp function, which is given by
\be\label{warpfunctionr}
A_r(y)=\frac{1}{9r}\left[(1-3r)\tanh^2(2ry)-2\ln\cosh(2ry)\right].
\ee
It is used to give the warp factor, which is depicted in Fig.~\ref{fig6}. We note that the warp function narrows as $r$ increases.

\begin{figure}[htb!]
\includegraphics[width=5.8cm]{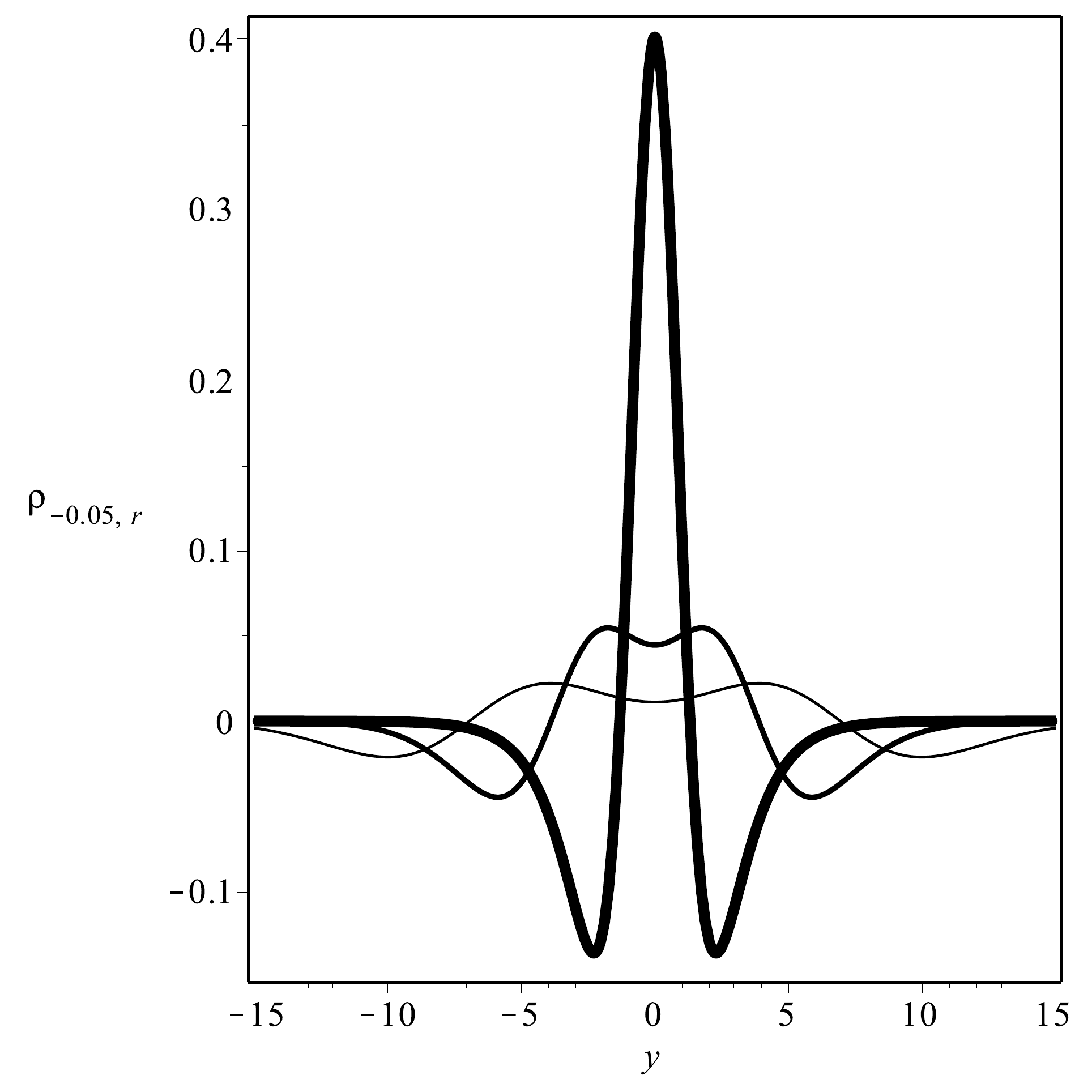}
\includegraphics[width=5.8cm]{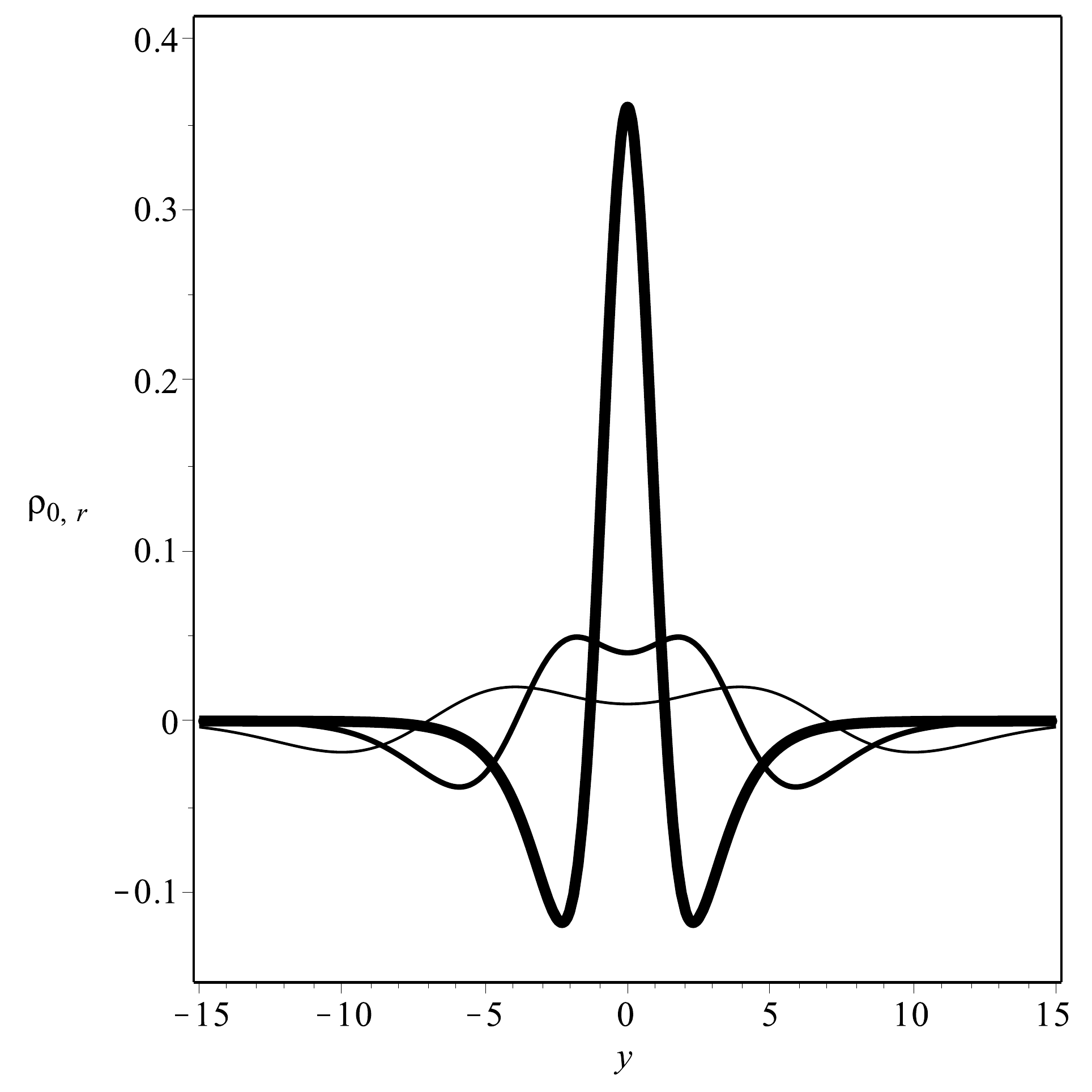}
\includegraphics[width=5.8cm]{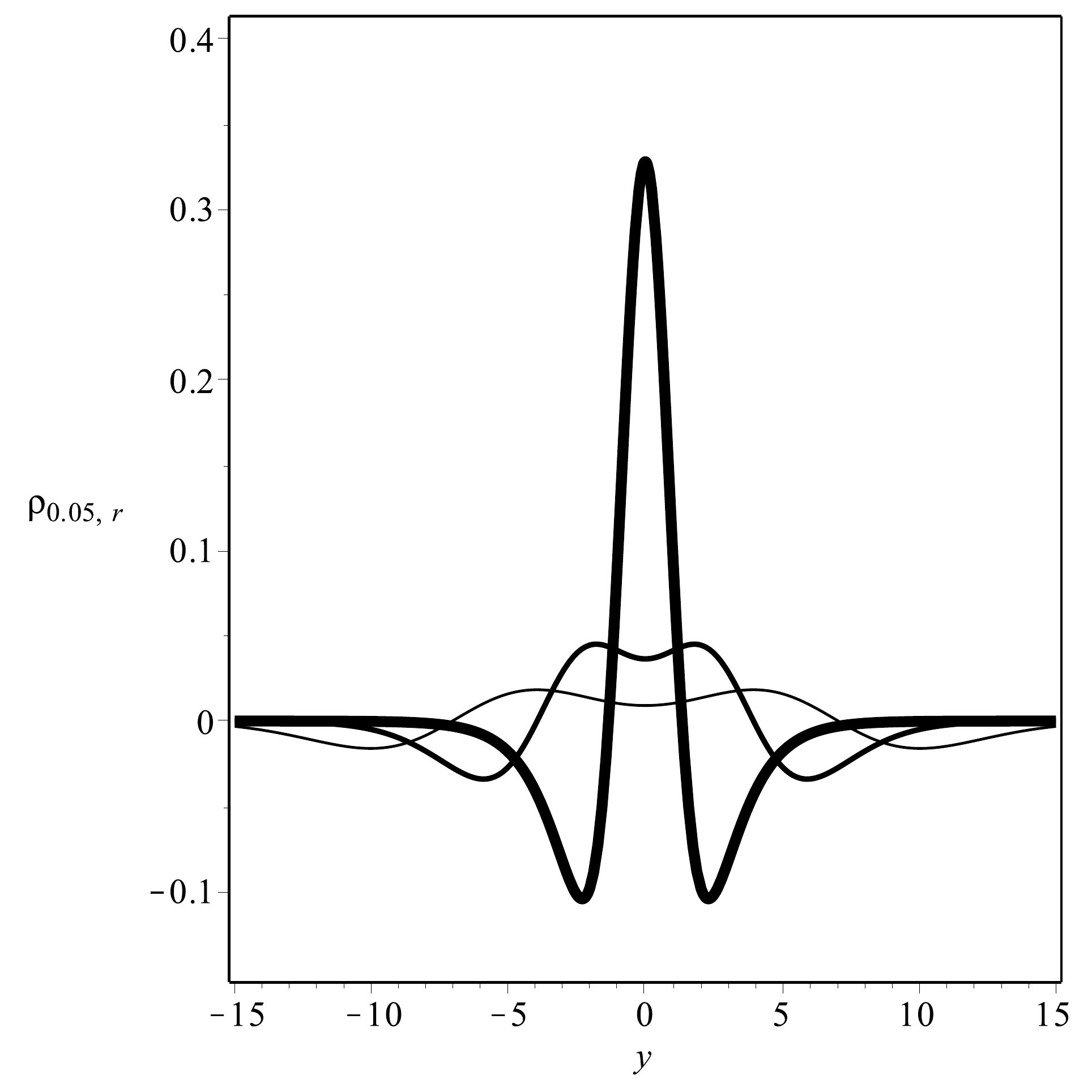}
\caption{The energy density $\rho_{\alpha,r}$ depicted for $\alpha=-0.05$ (left), $\alpha=0$ (center) and $\alpha=0.05$ (right). We have used $r=0.05, 0.1$ and $0.3$, and the thickness of the lines grows as $r$ increases.}
\label{fig7}
\end{figure}

Also, we can substitute Eq.~\eqref{superpotentialbnrt} into Eq.~\eqref{rho} and combine it with Eqs.~\eqref{solphir}, \eqref{solchir} and \eqref{warpfunctionr} to find the energy density $\rho_{\alpha,r}$, which is plotted in Fig.~\ref{fig7}. Here we also note that although there is quantitative modification in the braneworld profile, no new qualitative effect is induced by the minimal modification that we are considering in the current work. In particular, we note from Fig.~\ref{fig7} that the brane starts to split as $r$ varies from $r=0.3$ to lower and lower values, although the splitting is almost insensitive to small variations in $\alpha$.

\subsection{The hybrid brane model}

\begin{figure}[htb!]
\includegraphics[width=5cm]{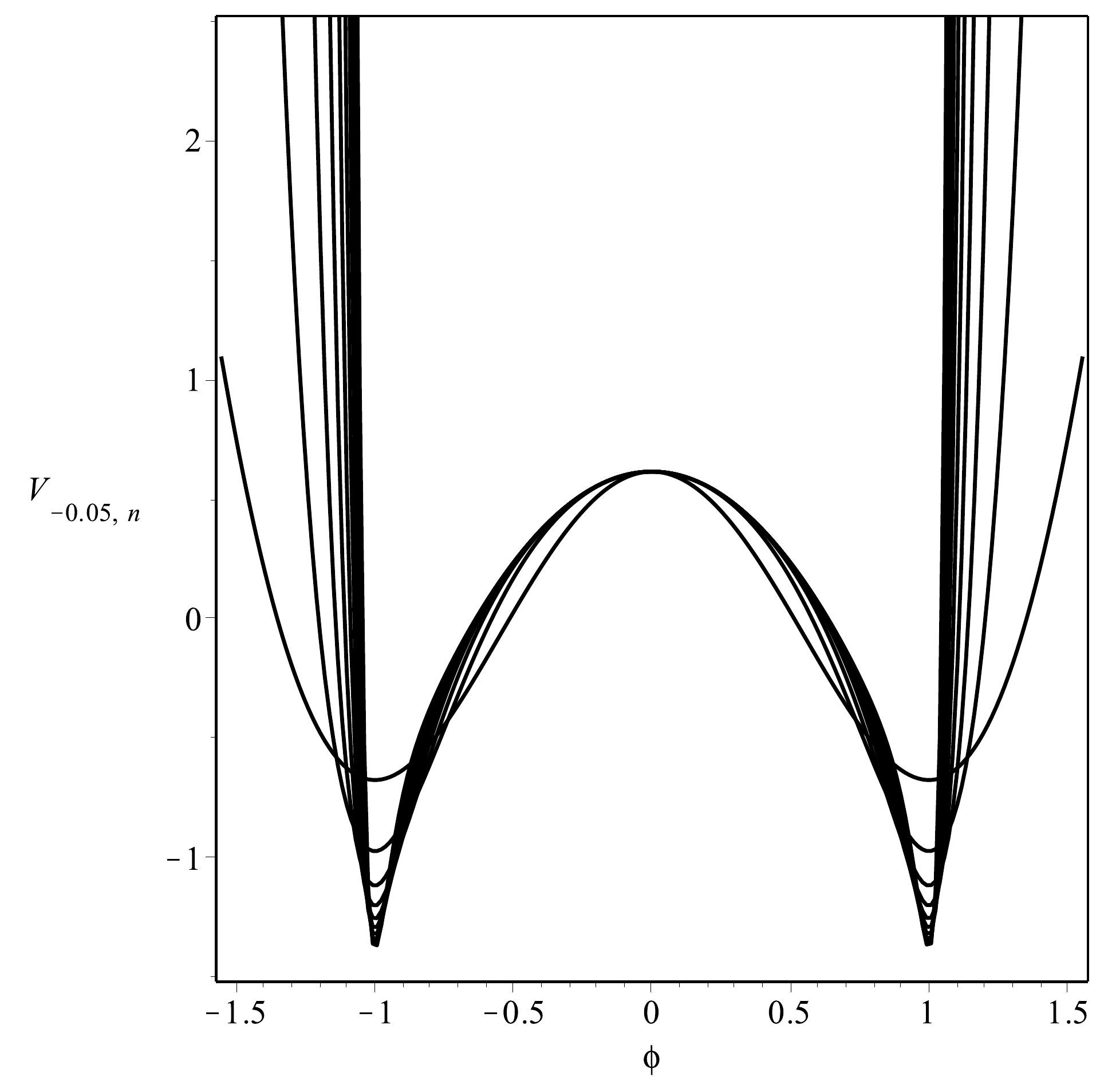}
\hspace{1cm}
\includegraphics[width=5cm]{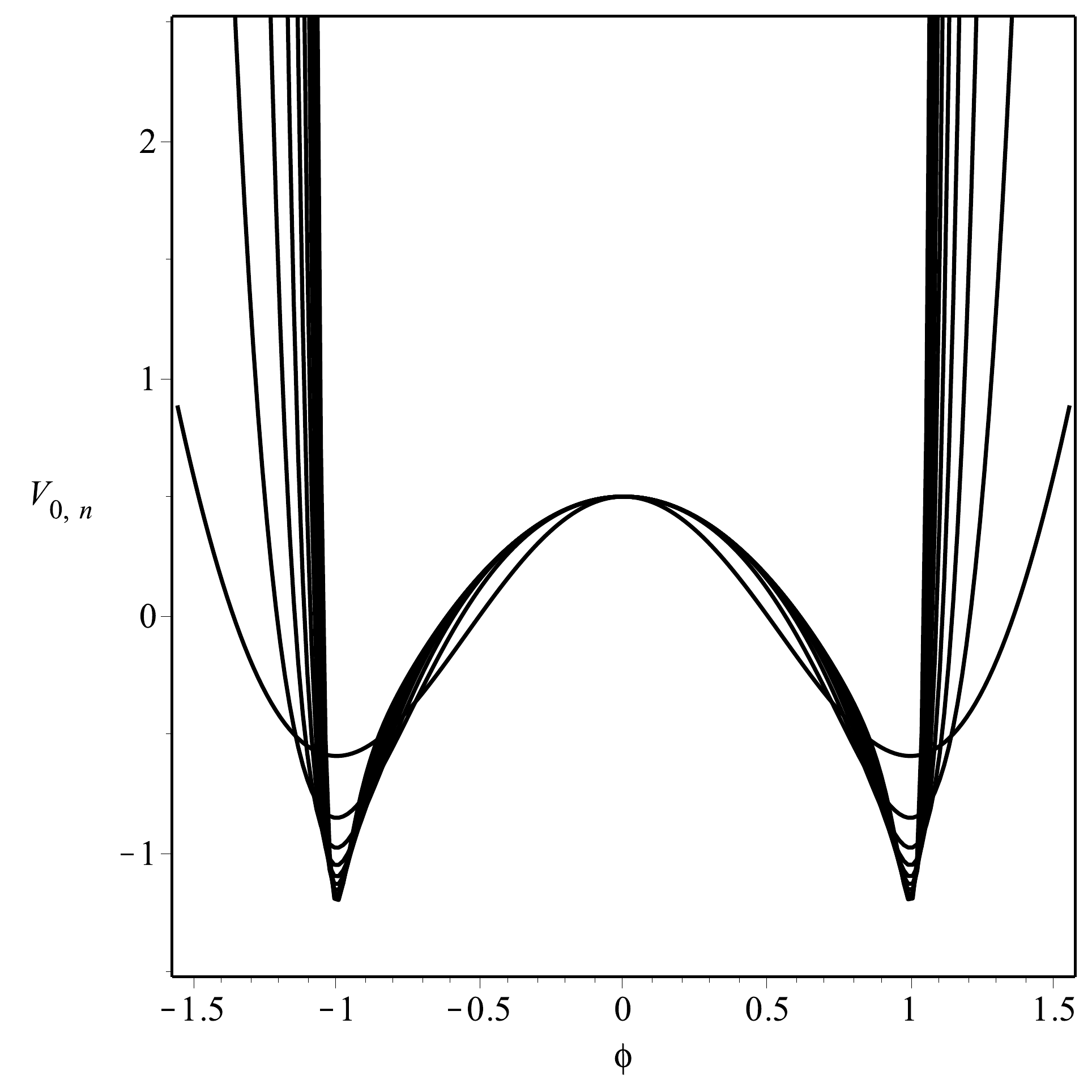}
\hspace{1cm}
\includegraphics[width=5cm]{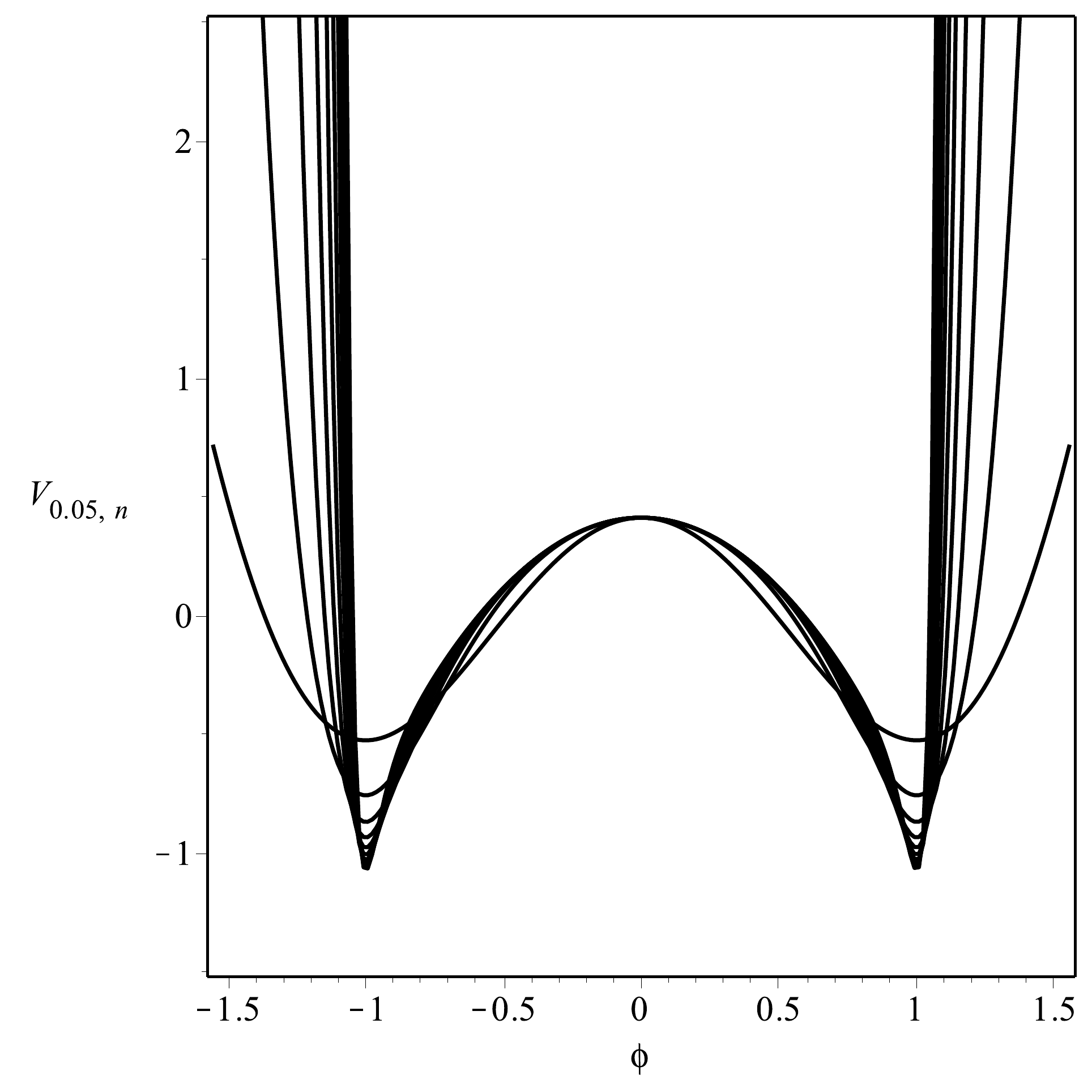}
\caption{The potential $V_{\alpha,n}$ depicted for $\alpha=-0.05$ (left), $\alpha=0$ (center) and $\alpha=0.05$ (right). Also, we have used $n=1,2,\ldots,10$.}
\label{fig8}
\end{figure}

Let us now investigate the model recently introduced in \cite{blmm}. It is defined by the following superpotential for a single field $\phi$
\be
W_n(\phi) = 2\phi - \frac{2\phi^{2n+1}}{2n+1}.
\ee
Here, $n$ is a positive integer parameter. This model was investigated in \cite{blmm}, and there is was shown that the brane has hybrid behavior, being thin or thick, depending on the extra dimension being inside or outside a compact space. The point here is that for $n$ increasing to larger and larger values, if the extra dimension varies in the compact interval $[-1,1]$, the warp factor has a thick profile. However, if the extra dimension varies outside the compact interval, it behaves as a thin brane. This makes the brane hybrid, as we further comment below Eq.~(28) and illustrate in Fig.~\ref{fig10}. To see how the hybrid scenario changes in the present context, here we note that the potential has the form
\be
V_{\alpha,n}(\phi) = \frac{2-3\alpha}{2(2+5\alpha)}\left(1-\phi^{2n}\right)^2- \frac{8}{3(2+5\alpha)}\left( \phi - \frac{\phi^{2n+1}}{2n+1} \right)^2.
\ee
It is depicted in Fig.~\ref{fig8} for several values of $\alpha$ and $n$.

\begin{figure}[htb!]
\includegraphics[width=6cm]{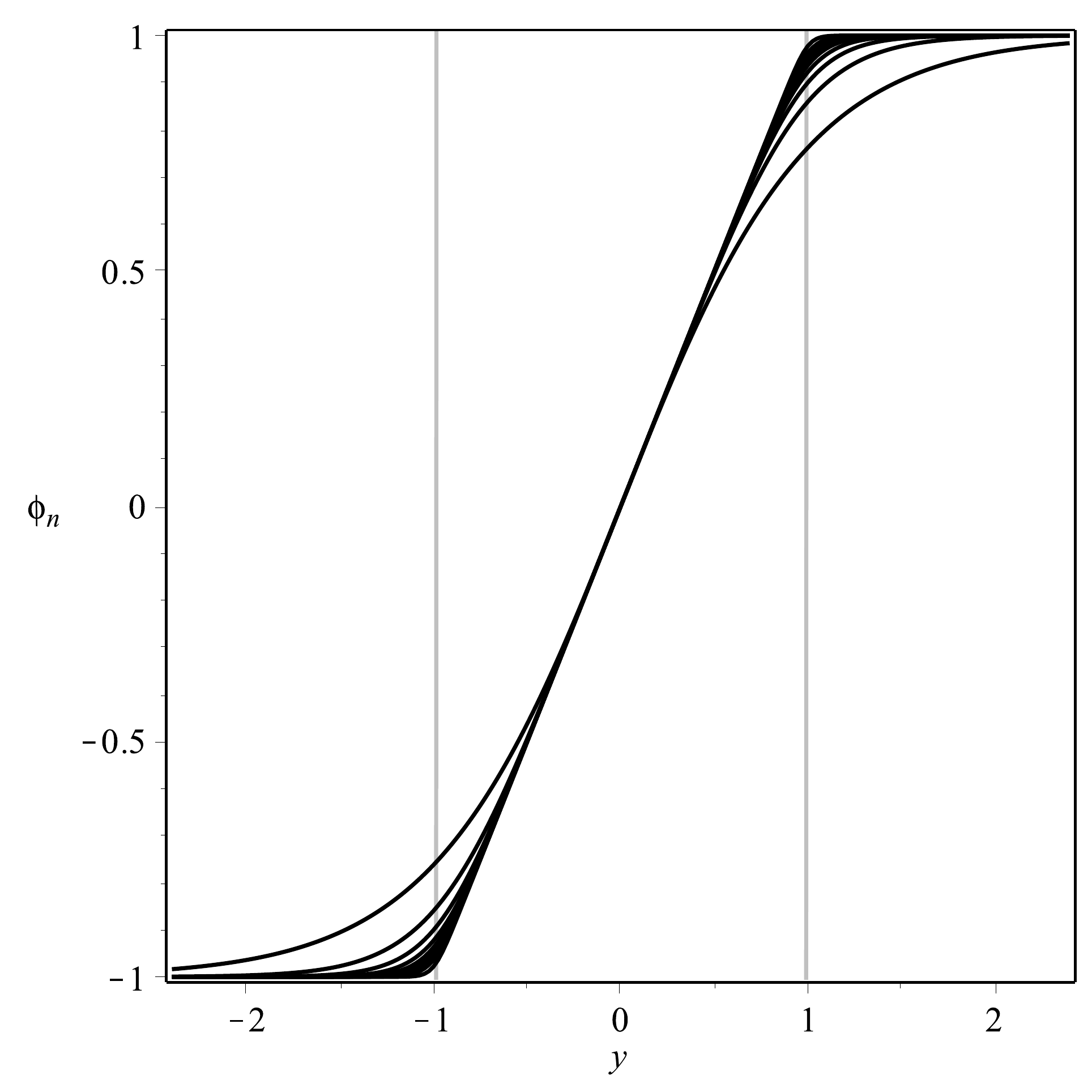}
\caption{The solution of the Eq.~\eqref{fohbphi} depicted for several values of $n$.}
\label{fig9}
\end{figure}

\begin{figure}[htb!]
\includegraphics[width=6cm]{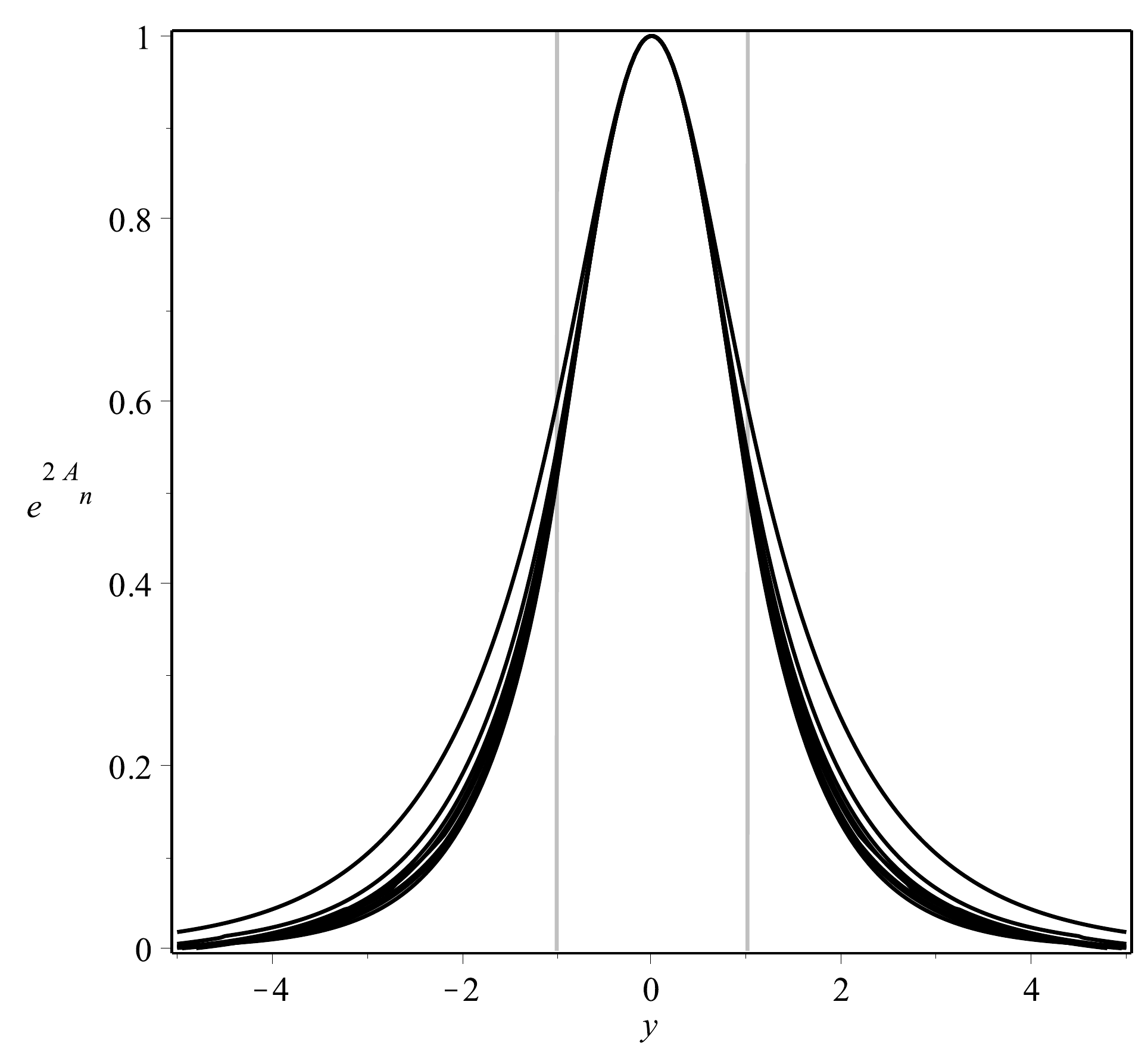}
\caption{The warp factor for the warp function \eqref{warpfunctionhb} depicted for several values of $n$.}
\label{fig10}
\end{figure}

The equation \eqref{firstorderb} becomes
\be\label{fohbphi}
\phi^\prime = 1-\phi^{2n}.
\ee
We have plotted its solution for several values of $n$ in Fig.~\ref{fig9}.

For $n$ very large, the solution tends to a compact kink
\begin{equation}\label{ckn}
\phi_c(x)=\left\{
\begin{array}{ll}
1, &  \mbox{ for } x> 1;\\ 
x,& \mbox{ for } |x|\leq 1;\\ 
-1,  & \mbox{ for } x < -1.
\end{array} \right.
\end{equation}

Moreover, we can use both first-order equations \eqref{firstorderb} and \eqref{firstordera} to write the warp function in terms of the scalar field analytically, in the form 
\be\label{warpfunctionhb}
A_n(y)=-\frac13{\frac {{[\phi(y)]}^{2}}{2\,n+1}}-\frac{2n{[\phi(y)]}^{2}}3{\frac {\,
{\,\mbox{$_2$F$_1$}(1,\frac{1}{n};1\!+\!\frac{1}{n};\!{[\phi(y)]^{2n}})}}{2\,n+1}},
\ee
where $_2F_1$ is hypergeometric function. In Fig.~\ref{fig10} we used this warp function to depict the warp factor for several values of $n$. It is worth pointing out that if $n$ is very large, it can be easily seen by using eq.~\eqref{firstordera} that the warp function tends to assume the form
\be
A_c(y)=
\begin{cases}
-\frac13 y^2,\,\,\,&|y|\leq 1;\\
-\frac23 |y|+\frac13, \,\,\, & |y|>1.
\end{cases}
\ee 
We see very clearly that for $y$ inside the compact interval $[-1,1],$ the warp factor behaves as $\exp(-y^2/3)$, in a way similar to a thick brane, and for $y$ outside the interval, the warp factor behaves as $\exp{(-2|y|/3)}$, so it engenders the standard thin brane behavior.

\begin{figure}[htb!]
\includegraphics[width=5cm]{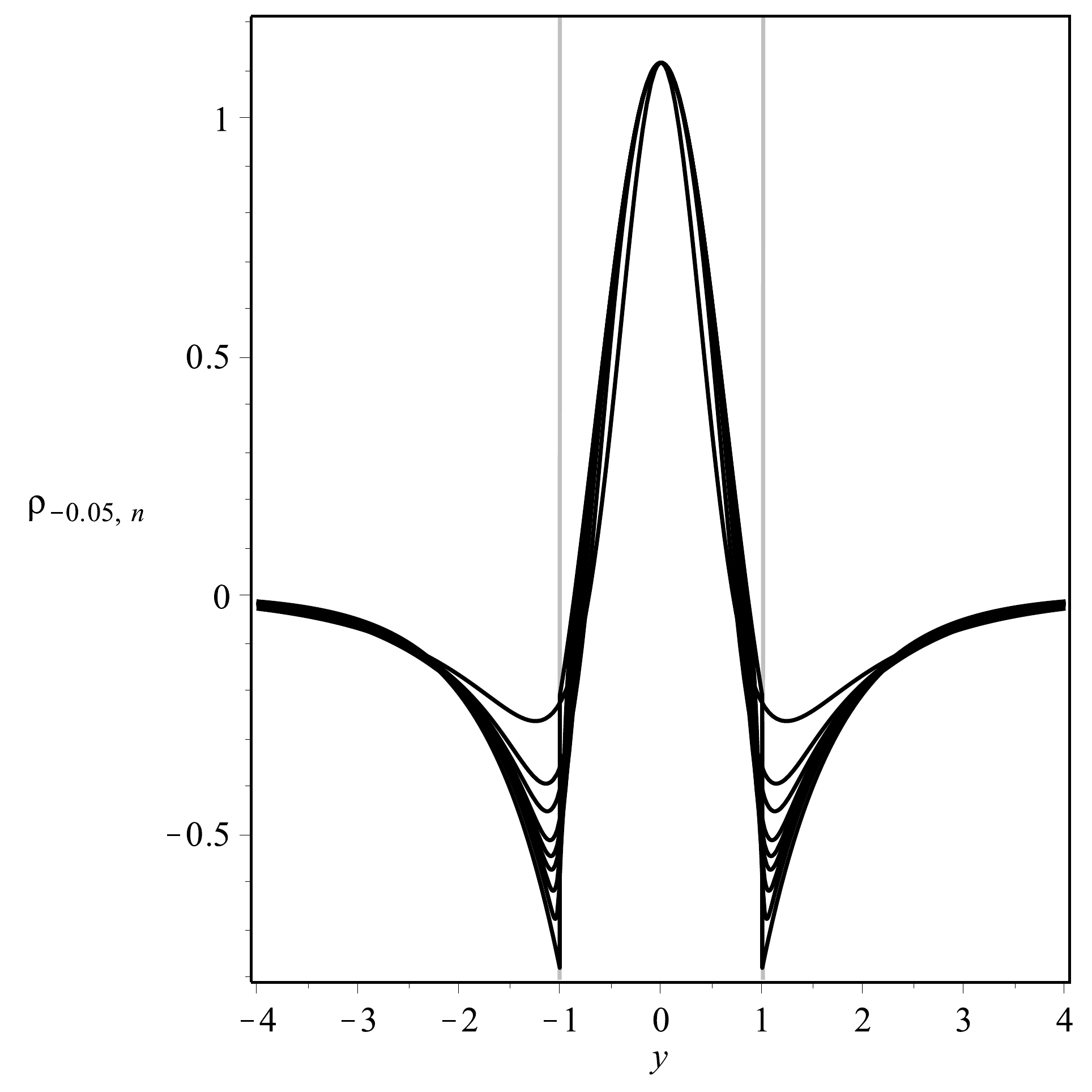}
\hspace{1cm}
\includegraphics[width=5cm]{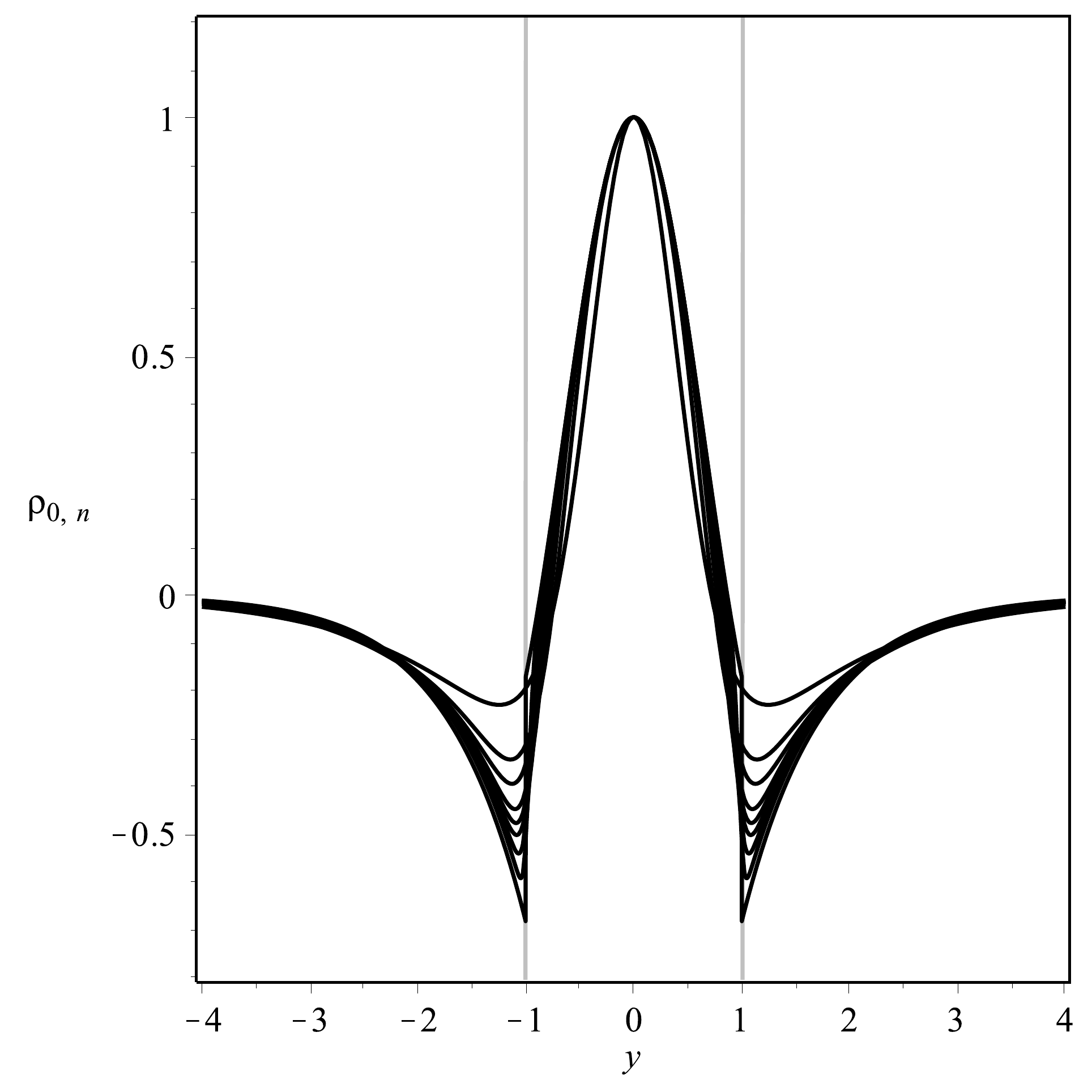}
\hspace{1cm}
\includegraphics[width=5cm]{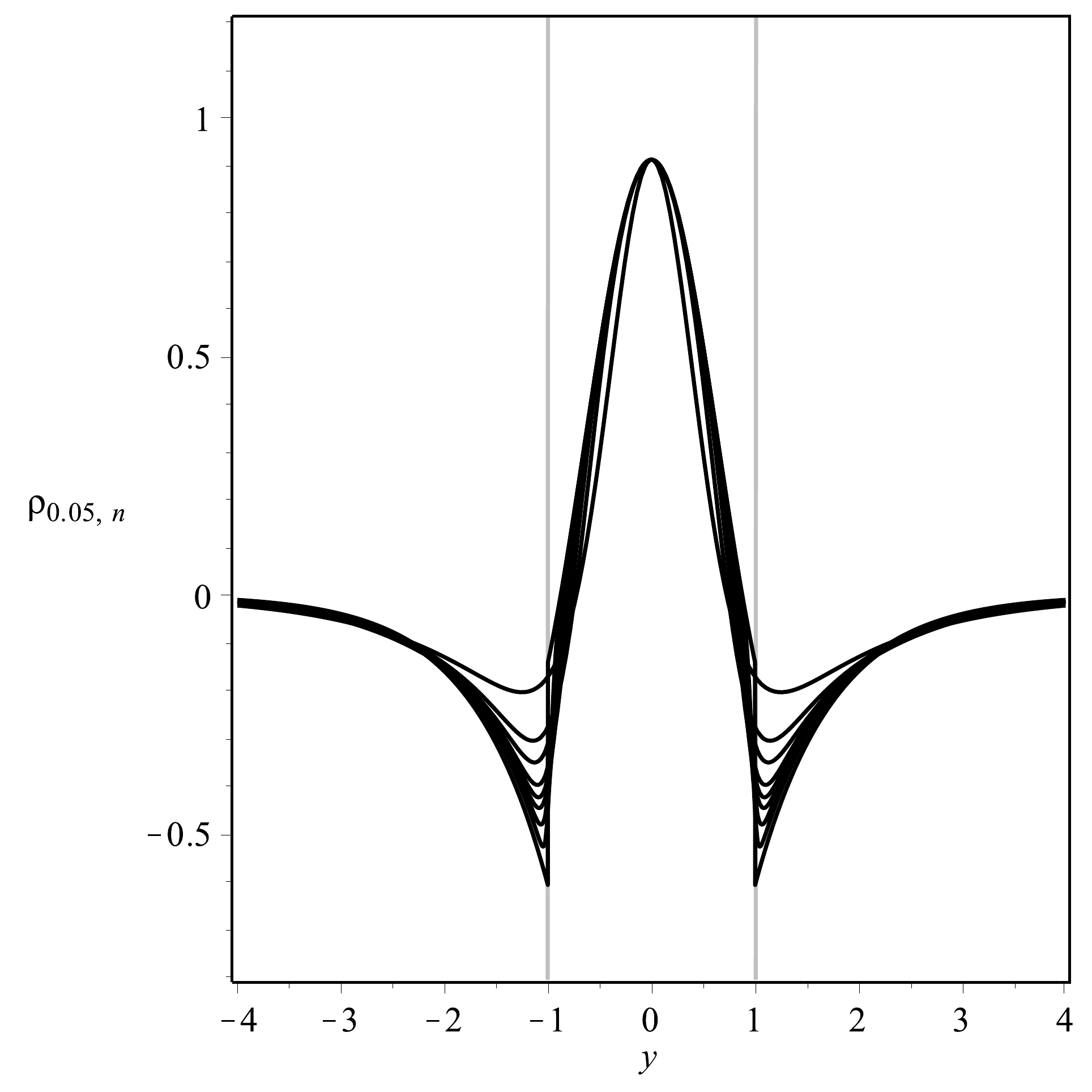}
\caption{The energy density $\rho_{\alpha,n}$ depicted for $\alpha=-0.05$ (left), $\alpha=0$ (center) and $\alpha=0.05$ (right) for several values of $n$.}
\label{fig11}
\end{figure}

Finally, the energy density $\rho_{\alpha,n}$ is depicted in Fig.~\ref{fig11}. The process is done numerically. Nevertheless, in the compact limit, for $n\to\infty$ we reach the following analytical result
\be
{\rho^{c}}_{\alpha,n}(y)=e^{2A_c(y)}
\begin{cases}
-\frac{8}{3(2+5\alpha)} y^2 + \frac{2+\alpha}{2+5\alpha},\,\,\,&|y|\leq 1,\\
-\frac{8}{3(2+5\alpha)}, \,\,\, & |y|> 1.
\end{cases}
\ee

We see from the energy density depicted in Fig.~\ref{fig11} that although there is quantitative modification, no new qualitative effect is induced by the minimal modification here considered. However, we note that the height of the maxima at zero and the depth of the minima depend on the value of $\alpha$, and this may induce modifications of phenomenological interest \cite{AH,KOP,blmm}.

\section{Comments and conclusions}
\label{sec:end}
In this work we studied braneworld models in the presence of auxiliary fields. The models that we investigated were described in an $AdS_5$ environment, with the presence of a single extra spatial dimension of infinite extent. 

We first extended former results to the case of several real scalar fields, and then we investigated three distinct models, one in the case of a single field, but with self-interaction controlled by an odd integer, which introduced the interesting effect of splitting the brane \cite{p}, another one described by two fields, giving rise to the Bloch brane scenario \cite{bg}, and the last one, which gives rise to hybrid brane \cite{blmm}. We could see that the presence of $\alpha$ induce modifications in each one of the braneworld scenarios, but the modifications do not qualitatively change the corresponding braneworld scenarios. These results add to the one obtained in \cite{we}, where it was also shown that the presence of $\alpha$ induces quantitative modification, without changing qualitatively the braneworld scenario of the models there investigated.

Based on the two models investigated in Ref.~\cite{we} and in the three models studied in the present work, we then conjecture that the minimal modification of Einstein's equation, controlled by $\alpha$ in
\be
R_{ab} - \frac12 g_{ab}R = 2T_{ab} + \alpha g_{ab}T,\nonumber
\ee
where $T$ is the trace of the energy-momentum tensor $T_{ab}$, is not able to produce qualitative modifications in the braneworld scenario
sourced by real scalar fields in the case of vanishing $\alpha$. In this way, to further probe the braneworld scenario in the presence of auxiliary fields we should then add more contributions, going beyond the minimal modification that we have studied, as suggested in the original work \cite{psv}. Another issue of interest concerns the first-order framework set forward in Ref.~\cite{we} and here extended to several fields; it suggests the presence of supersymmetry, so one could ask how the presence of auxiliary fields complies with supersymmetry.

A natural extension of the current work concerns the addition of other terms to the source Lagrange density, having higher-order power in the first derivative of the field, as a way to change kinematics of the scalar field, to see how it can comply with the auxiliary field framework here considered. The basic steps toward this line of investigation was already given in the recent work \cite{epjc}. Another extension of current interest concerns the addition of fermion and gauge fields, to investigate how the quantitative modifications here unveiled contribute to control the trapping of those fields inside the brane.

\acknowledgements

The authors would like to thank CAPES and CNPq for partial financial support.


\end{document}